\documentclass{iopjournal}
\usepackage[version=3]{mhchem} 
\usepackage{lmodern}

\newcommand{\kjmol}{kj\,mol$^\text{-1}$}

\begin{document}

\articletype{Paper} 

\title{Enhancing Hydrogen Adsorption Ability of MOF-5 with Metal Node Exchange and Linker Functionalisation}

\author{
 Joshua Edzards$^{1,2}$
 Holger-Dietrich Saßnick$^{1,3}$
 Caterina Cocchi$^{1,2,4}$
}

\affil{$^1$Carl von Ossietzky Universit\"at Oldenburg, Institute of Physics, 26129 Oldenburg, Germany}

\affil{$^2$Friedrich-Schiller Universit\"at Jena, Institute for Condensed Matter Theory and Optics, 07743 Jena, Germany}

\affil{$^3$ICGM, Université de Montpellier, CNRS, ENSCM, 34293 Montpellier, France}

\affil{$^4$Abbe Center of Photonics, Friedrich-Schiller-Universit\"at Jena, 07745, Jena, Germany}

\email{joshua.edzards@uni-oldenburg.de, caterina.cocchi@uni-jena.de}



\begin{abstract}
In the search for promising material candidates for hydrogen storage, we investigate from first principles derivatives of metal organic framework 5 (MOF-5), including isoelectronic substitution of the metal centers (Zn $\rightarrow$ Mg, Cd) and linker functionalization with \ce{NH2}, OH, and \ce{NO2} groups.  Metal-node substitution consistently stabilises, and ligand functionalization systematically enhances \ce{H2} binding at the metal-oxo cluster sites. The combination of Cd centers and \ce{NO2} groups proves to be most efficient, yielding an adsorption energy of -15.58~kJ mol$^{\text{-1}}$, which substantially outperforms the storage ability of pristine MOF-5. Detailed electronic structure analysis clarifies how the local framework environment coordinates the guest molecule, providing a robust design framework to guide the experimental development of advanced adsorbent materials.
\end{abstract}

\section{Introduction}

Global transition toward sustainable energy hinges on the development of efficient, safe, and scalable hydrogen storage technologies~\cite{ma+24ijhe,xu+24ef}. As a zero-emission fuel, hydrogen holds immense promise for decarbonizing transportation and energy sectors, particularly through fuel cell applications~\cite{staf+19ees}. However, the practical deployment of hydrogen-powered systems is severely limited by the challenges associated with their storage~\cite{ren+15ijer}. Conventional methods, such as high-pressure compression (up to 700~bar) or cryogenic liquefaction, require energy-intensive processes and robust, heavy containment systems, undermining the overall efficiency and economic viability of hydrogen as a fuel~\cite{riva+19mat}.

Metal-organic frameworks (MOFs) have emerged as a highly promising class of porous materials for solid-state hydrogen storage due to their exceptional surface areas, tuneable pore architectures, and chemical versatility~\cite{suh+12cr,zhan+23ica,qiao+24ijhe,shi+24ijhe,sutt+24cej}. Among them, MOF-5 (also known as IRMOF-1)~\cite{li+99nat,kaye+07jacs}, a prototypical isoreticular framework based on Zn oxo-clusters and 1,4-benzenedicarboxylate (BDC) linkers, has been extensively studied for its high porosity and structural stability~\cite{liu+08jacs,chen+10jmc,klei+10tha,hugh+11jacs,yang+12ijhe,ming+16ijhe}. Despite these advantages, pristine MOF-5 exhibits relatively weak hydrogen storage capacity~\cite{cabr+08prb}, with adsorption energies typically around (-4.6 to -7.6)~kJ mol$^{\text{-1}}$~\cite{saga+04jcp,muel-cede05jpcb,dail+06jpcb,fitz+08prb,gall-glos09jpcc,sill+09jacs}, which fall short of the ideal range (-15 to -25~kJ mol$^{\text{-1}}$) required for reversible, near-ambient hydrogen storage with sufficient gravimetric and volumetric capacity~\cite{sutt+24cej,medu-nand23mtp}.

To overcome this limitation, extensive efforts have focused on modulating MOF-5 through structural and chemical engineering~\cite{yan+23acsami,meek+24natm,orel24jpcc,edza+24jcp} to enhance the interaction with hydrogen molecules. While previous studies have explored individual modifications, a systematic, high-throughput investigation of the combined effects of metal node variation and functional group incorporation is still missing. Likewise, beyond data generation, a profound understanding of the underlying electronic mechanisms, such as charge transfer, polarization, and orbital hybridization, is essential to guide rational design rather than empirical screening.

In this work, we present a comprehensive first-principles study of \ce{H2} adsorption in a set of MOF-5 variants, hereafter termed M-MOF-5-X, where M indicates the metal node (Zn, Mg, Cd), and X the functional group (\ce{NH2}, OH, \ce{NO2}) at the 2,5-positions of the BDC linker. Using density-functional theory (DFT) with the R2SCAN-D3 functional and counterpoise corrections, we evaluate adsorption across three distinct framework positions. Sites near the metal-oxo clusters exhibit the highest binding affinities, confirming trends observed in earlier studies~\cite{muel+05jpc, sill+09jacs}. To manage this screening efficiently, we integrated our computational workflows into the open-source \texttt{aim$^2$dat} library, ensuring consistency and reproducibility. We identify Cd-MOF-5-\ce{NO2} as the most promising candidate, exhibiting the strongest adsorption energy in the small pore, significantly outperforming pristine MOF-5 and all other variants. Electronic structure analysis reveals that this enhancement stems from the synergistic interaction between the Cd node and electron-withdrawing \ce{NO2} groups within the framework's small cage. Overall, this study identifies a promising candidate for hydrogen storage while offering a transferable, high-throughput computational framework for the rational design of functionalized porous materials.

\section{Methods and Systems}

\subsection{Workflow and Computational Details\label{sec:comp_details}}
The computational analysis presented in this study is executed using a custom automated workflow integrated into the \texttt{Python} library \texttt{aim$^\text{2}$dat}~\cite{sass+26es} (further details in Section~S1 of the Supplementary Information), with provenance and data management handled through the \texttt{AiiDA} ecosystem~\cite{hube+20sd,uhri+21cms}. Symmetry-inequivalent adsorption sites are identified using \texttt{spglib}~\cite{togo+24stamm} to ensure systematic placement of guest molecules. Density functional theory (DFT) calculations are performed in sequential steps: initial self-consistent field (SCF) parameter convergence, optimization of the guest molecules using progressively tighter convergence thresholds, and final electronic structure evaluation. Adsorption energies are corrected for basis set superposition error (BSSE) via the counterpoise method, and projected density of states (pDOS) are generated to analyze host-guest electronic interactions on an atomistic, quantum-mechanical level. Post-processing and visualization are performed with the routines available in \texttt{aim$^\text{2}$dat}~\cite{sass+26es}.

DFT calculations~\cite{hohn+64pr,kohn+65pr} were performed using \texttt{CP2K}~\cite{kueh+20jcp} (version 2025.1) adopting Goedecker-Teter-Hutter pseudopotentials~\cite{goed+96prb} and MOLOPT Gaussian basis sets~\cite{vand+07jcp} optimised using the UZH protocol. Based on a previous benchmark study~\cite{edza+25jctc}, the meta-GGA functional R2SCAN~\cite{furn+20jpcl} including the Grimme-D3~\cite{grim+10jcp} correction was adopted. For the final configurations, a 4$\times$4$\times$4 Monkhorst-Pack~\cite{monk+76prb} \textbf{k}-mesh was employed alongside a triple-zeta valence double polarization basis set, with plane-wave and relative cutoff energies of 1500~Ry and 200~Ry, respectively. The pre-optimization phase was conducted with looser settings, including a double-zeta valence single-polarization basis set with a reduced cutoff of 1000~Ry, subject to pressure and force convergence thresholds of 200~bar and 0.005~Hartree/Bohr, respectively.  Finally, BSSE were corrected using the counterpoise method~\cite{boys+70mp,krus+12jcp} to ensure accurate adsorption energy evaluations.

\subsection{Construction of the Systems\label{sec:systems}}

Pristine MOF-5 forms a cubic framework ($Fm\overline{3}m$ space group, lattice constant $a = 25.87$~\AA)~\cite{li+99nat} composed of inorganic $\text{Zn}_4\text{O}$ oxo-clusters connected by 1,4-benzenedicarboxylate (BDC) linkers. In previous work, we systematically investigated the effects of metal-node exchange and linker functionalization on the energetic, structural, electronic, and vibrational properties of the framework~{\cite{edza+24jcp}, substituting Zn metal centers with isoelectronic elements such as Mg and Cd yields tuned geometries and electronic properties due to varying atomic radii and screening effects~\cite{wen+21cell}. Furthermore, functionalizing each BDC phenyl ring at two hydrogen positions with electron-donating groups like \ce{NH2} and OH stabilizes hydrogen-bonding networks with minimal backbone distortion, yielding smaller electronic bandgaps via localized states at the valence band maximum. Conversely, the strongly electron-withdrawing \ce{NO2} group induces steric repulsion and linker rotations, modifying local charges and contributing directly to both band edges~\cite{edza+24jcp}.
Phonon dispersion calculations and detailed formation energy analysis presented in previous work~\cite{edza+24jcp,sant-cocc26es} confirm the dynamic and energetic stability of the considered MOF-5 derivatives. 

\begin{figure}
\begin{center}
\includegraphics[width=\textwidth]{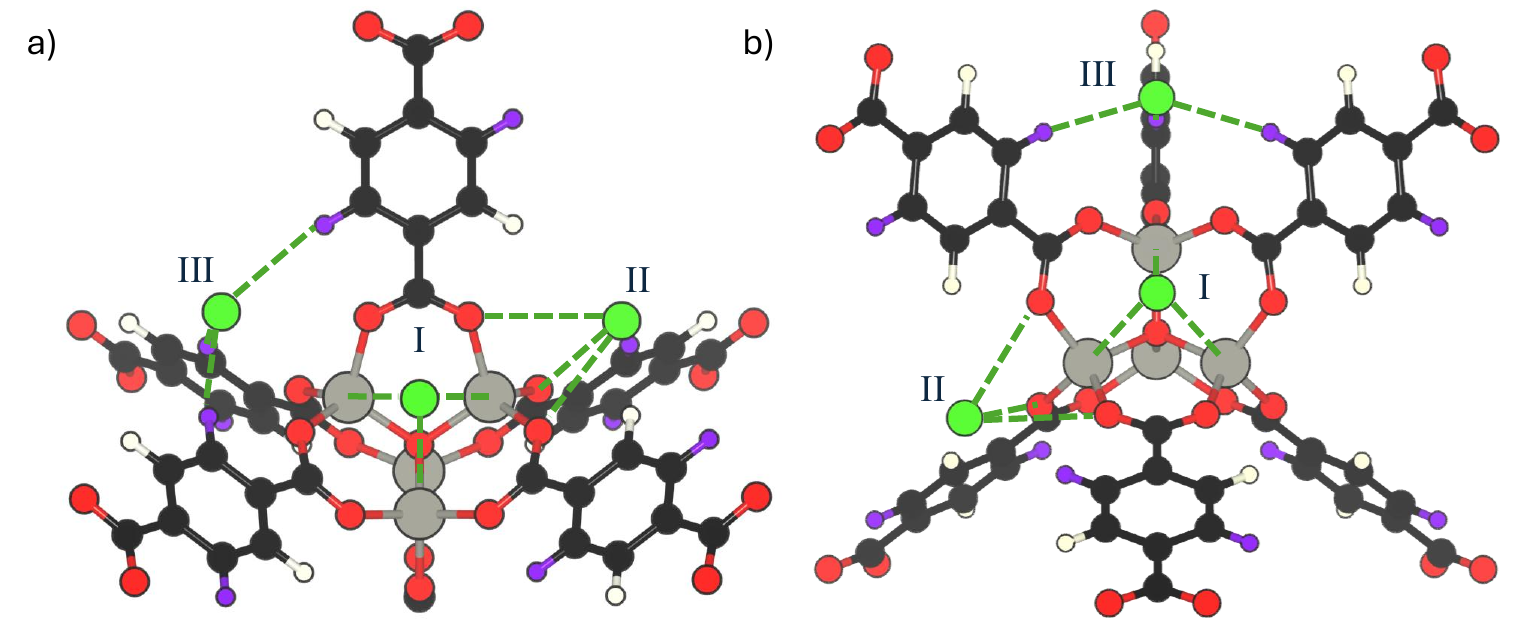}
\caption{\label{fig:H2_positions} Structural representations of the host framework viewed from different perspectives, a) and b), highlighting the three distinct \ce{H2} adsorption environments: (I) large-pore metal site, (II) small-pore oxygen-anchored site, and (III) small-pore site enclosed by three functional groups. Metal nodes are depicted in grey, oxygen atoms in red, carbon atoms in black, passivating hydrogens in white, and functional groups (violet). The adsorbed \ce{H2} molecule is represented by a bright green sphere, with green dashed lines indicating the primary anchor points for guest–host interaction.
}
\end{center}
\end{figure}

Based on symmetry analysis, we identify three inequivalent adsorption sites for molecular hydrogen (Figure~\ref{fig:H2_positions}a-b). Position I is situated in the large pore of the framework adjacent to the $\text{M}_4\text{O}$ cluster, where three metal cations serve as primary anchor points. The second site (II) is positioned in the small pore near the oxo-metal cluster, anchored by three carboxylate oxygen atoms in proximity to at most one functional group in the linker. The third adsorption position (III) is also in the small pore close to the oxo-metal cluster, but directly enclosed within the steric and electronic influence of three converging functional groups.

\section{Results}

\subsection{Structural Properties\label{sec:strct_prop}}

\begin{figure}
\begin{center}
\includegraphics[width=\textwidth]{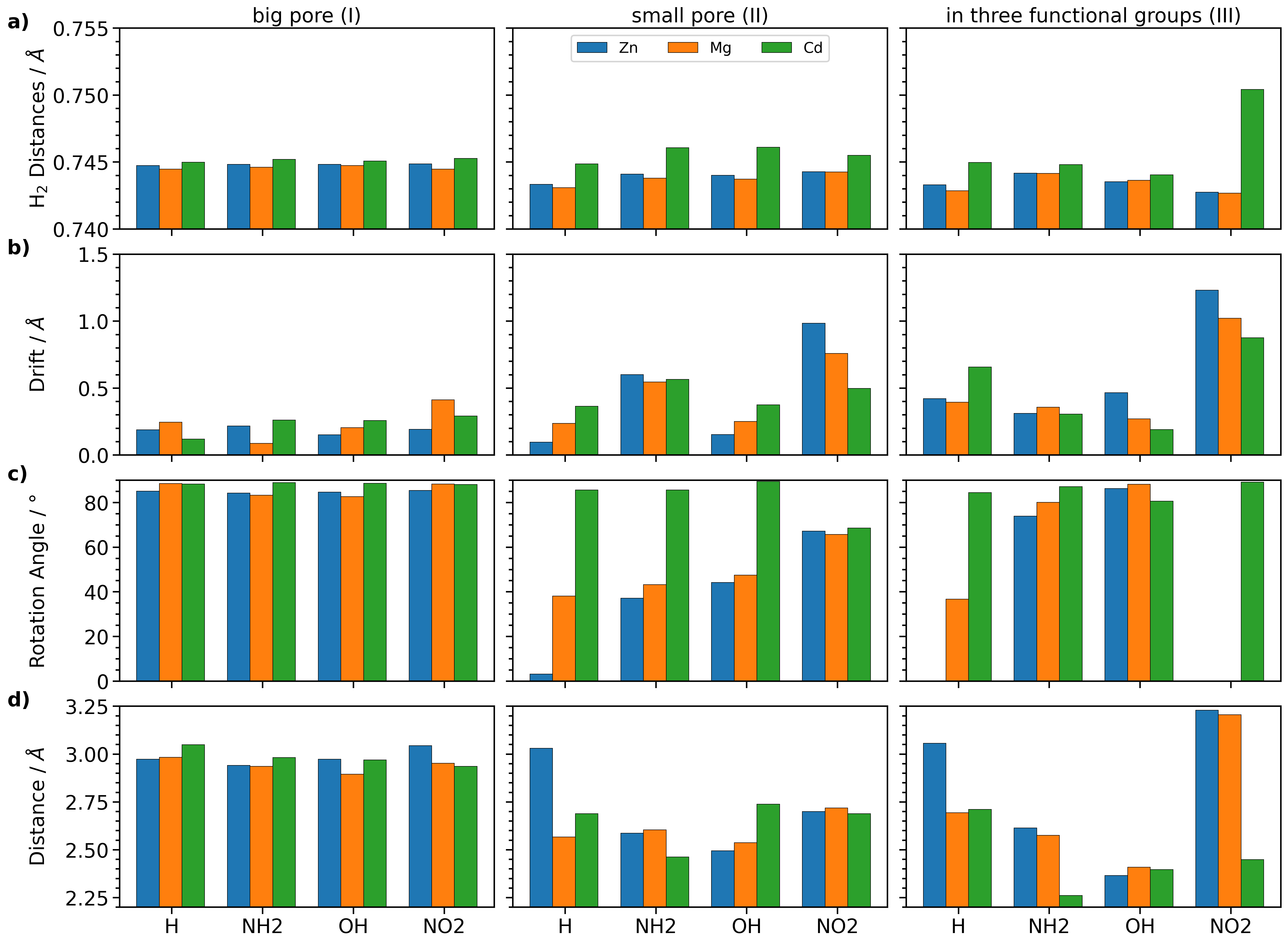}
\caption{\label{fig:structural_properties} a) Intramolecular H–H bond distance, b) center-of-mass drift from the initial symmetry site, c) molecular rotation angle relative to the intial orientation, and d) the shortest host–guest distance of the considered M-MOF-5-X derivatives across three distinct chemical environments: (I) large pore, (II) small pore, and (III) small pore enclosed by three functional groups. 
}
\end{center}
\end{figure}

The structural properties of the MOF-5 variants follow the trends established in earlier studies~\cite{edza+24jcp, edza+25jctc}. Consequently, the analysis herein focuses directly on the structural perturbations and local coordination of the adsorbed \ce{H2} molecule. Detailed results are provided in the Supplementary Material, Tables~S1--S4.
We first examine the H-H bond length (Fig.~\ref{fig:structural_properties}a). For all investigated frameworks and adsorption sites, the relaxed H–H distance exceeds the experimental gas-phase value of 0.741~\AA{}~\cite{hube+13springer}. In pristine MOF-5 (with Zn metal center), relaxation yields an H-H bond length of 0.745~\AA{} at the large-pore metal site (I) and 0.743~\AA{} at both small-pore sites (II and III). These values are in excellent agreement with prior semi-local DFT calculations (0.745~\AA{})~\cite{muel+05jpc} and quantum-chemical analyses on benzoate cluster models (0.741~\AA{} and 0.739~\AA{} for sites I and II, respectively)~\cite{sill+09jacs}. This systematic bond elongation upon adsorption is corroborated by infrared spectroscopy data~\cite{fitz+08prb}, reporting a maximum vibrational redshift of $27\text{ cm}^{-1}$ at the primary $\text{Zn}_4\text{O}$ adsorption site, directly tying anharmonic bond stretching to local framework interactions.

Isoelectronic substitution of the metal node noticeably modulates this bond length. Replacing Zn with Mg results in a slight reduction of the H-H distance across all adsorption sites, with the bond shortening the most in the small-pore environments (II and III). Conversely, Cd substitution yields H-H separations comparable with those obtained in pristine MOF-5 in the large pore (I), but triggers an expansion within the small pore (II). Interestingly, in the sterically confined site III, the H–H distance in the Cd-substituted framework is shorter than at position II. These variations correlate with the trends in metal ionic radii ($\text{Mg}^{2+} < \text{Zn}^{2+} < \text{Cd}^{2+}$)~\cite{cord+08dt}, suggesting that the larger spatial requirement of the Cd node decreases the available volume within the small pore, thereby enhancing host–guest Pauli repulsion and altering the balance of electrostatic forces acting on the adsorbate.

The impact of linker functionalization depends strongly on the specific pore environment. In the large pore (I), the presence of functional groups has a negligible effect on the H–H distance due to the larger spatial separation between linker and adsorbate. On the other hand, within the small pore environments (II and III) functionalization consistently elongates the H–H bond relative to the unfunctionalized counterpart. At site II, the intramolecular distances remain relatively uniform regardless of the attached group, as the molecule possesses sufficient spatial degrees of freedom to rearrange itself and minimize unfavorable interactions. At site III, structural confinement pushes the guest molecule in proximity with the three ligands. Under these conditions, steric hindrance and electrostatic couplings dominate, leading to a subtle bond contraction relative to site II. Among the substituents, electron-donating \ce{NH2} and \ce{OH} groups induce the largest bond elongations at site II, whereas the electron-withdrawing \ce{NO2} group maintains values closer to pristine MOF-5.

When examining the combined effects of metal node exchange and ligand functionalization, the Mg-substituted frameworks closely mirror the geometric trends of their Zn counterparts across all considered descriptors. In contrast, Cd-containing MOF-5 variants exhibit peculiar structural behaviors. At the single-functionalized small pore site (II), they display a more pronounced expansion of the intramolecular H–H distance (Figure~\ref{fig:structural_properties}a). Notably, Cd-MOF-5-\ce{NO2} stands out as a clear structural outlier when confined within the multi-functionalized environment of site III. At this position, it induces the largest H-H elongation, shifting the intramolecular bond length to 0.751~\AA{}, a stretch of approximately 0.01~\AA{}~compared to pristine MOF-5. 


The spatial drift of the guest molecule, defined as the geometric displacement of the \ce{H2} center-of-mass from its initial symmetry-assigned coordinates to its fully relaxed position, is illustrated in Fig.~\ref{fig:structural_properties}b. For pristine MOF-5, the initial structural positioning is highly representative of the equilibrium state, yielding minor displacements of 0.19~\AA{} and 0.10~\AA{} at sites I and II, respectively. A larger drift of 0.42~\AA{} appears at site III, suggesting that when placed in the vicinity of three unfunctionalized linkers, the hydrogen molecule dynamically adjusts its position closer to the core of the small pore.

Isoelectronic substitution of the metal node preserves these low-drift characteristics in Mg-containing frameworks, which lead to a maximum drift of 0.39~\AA~at site III and minor deviations at sites I and II compared to their pristine, Zn-based counterpart. In contrast, Cd substitution induces a more pronounced site-dependent drift. While the large-pore environment (site I) remains highly stable with a minimal drift of 0.12~\AA, the displacements increase to 0.36~\AA{} at site II and 0.66~\AA{} at III. This behavior highlights how the larger ionic radius of $\text{Cd}^{2+}$ selectively restricts the available volume within the small-pore environment, forcing a structural re-accommodation of the guest molecule.

Functionalizing the BDC linkers yields no meaningful change in drift at the large-pore site (I) across all variants. On the other hand, compared to metal-node substitution, ligand functionalization has a stronger impact on the drift in the small-pore environments (site II and III). This is due to two concomitant effects: steric hindrance and electrostatic gradients. Consequently, enhanced host–guest interactions drive the adsorbate away from its initial coordinates. 
Within the small pore (site II), where only a single adjacent linker is modified, the \ce{NH2} and \ce{NO2} groups repel the guest molecule, inducing a noticeable outward displacement. Conversely, the hydroxyl group undergoes a localized rotation to point toward the framework oxygen atoms, minimizing its effective steric footprint within the pore and resulting in a drift profile nearly identical to the H-terminated reference. The strongly electronegative nitro group drives the largest displacement at site II. When the guest molecule is instead placed within the highly confined environment of three functional groups (site III), the \ce{NH2} group shows no additional impact relative to the baseline, while the \ce{OH} group increases the drift slightly. Crucially, the \ce{NO2} groups mirror their site II behavior, yielding the most substantial displacements in this subsection.

The combination of metal node substitution and ligand functionalisation lead to similar trends as obtained for pristine MOF-5, with the largest deviation of 0.26~\AA{} occurring between the Zn and Cd variants of the \ce{OH}-functionalized framework. The most relevant deviation appears with the nitro groups, where the computed drift exceeds 1~\AA{} in the Zn- and Mg-containing MOFs. Here, the steric and electrostatic repulsion from the three \ce{NO2} groups pushes the \ce{H2} molecule away from the metal-oxo cluster. For the Cd framework, this outward drift amounts to 0.88~\AA{}: rather than being entirely repelled, the guest molecule becomes sterically and electrostatically trapped in a localized energy minimum between the large $\text{Cd}_4\text{O}$ cluster and the surrounding nitro ligands, consistent with the anomalous bond elongation discussed in Fig.~\ref{fig:structural_properties}a.


The spatial reorientation of the guest molecule is evaluated via the calculated rotation angle $\theta$, defined relative to its initial symmetry-assigned alignment (Fig.~\ref{fig:structural_properties}c). In the ideal starting configuration, the \ce{H2} molecular axis is directed toward the central oxo-cluster oxygen atom at the large-pore site (I), and toward the metal cations at both small-pore sites (II and III). For pristine Zn-MOF-5, the initial placement requires negligible adjustment within the small pore, yielding a minimal rotation of 3.16$^\circ$ at site II and maintaining its baseline alignment at site III. In contrast, optimization at the large-pore metal site (I) drives a significant perpendicular reorientation, rotating the molecular axis by 85.1$^{\circ}$.

Isoelectronic metal substitution breaks this spatial degeneracy, prompting distinct, site-specific rotational profiles. Incorporating Mg reserves the near-orthogonal orientation at the large-pore site ($\theta = 88.52^{\circ}$) but induces intermediate rotations of $38.08^\circ$ and $36.74^\circ$ at sites II and III, respectively. In these configurations, the \ce{H2} molecule tilts away from the cation to align toward the carboxylate oxygen atoms of the BDC linker. Conversely, replacing Zn with the larger Cd atom exerts little impact on the large-pore site ($\theta = 88.92^\circ$) but forces a complete, near-orthogonal reorientation ($\theta \approx 90^\circ$) at both small-pore sites (II and III), underscoring the strong role of metal ionic radii in sterically steering the guest axis.

Ligand functionalization introduces an additional layer of directional control. Within the large-pore environment (site I), all functionalized variants uniformly drive a near-$90^\circ$ rotation, confirming that the long-range electrostatic field of the modified linkers stabilizes a perpendicular configuration relative to the oxo-cluster. At site II, the presence of a single functional group yields intermediate rotations of $37.13^\circ$, $44.20^\circ$, and $67.21^\circ$ for \ce{NH2}, \ce{OH}, and \ce{NO2}, respectively. In each case, the hydrogen atom most distant from the metal cluster points toward the substituent while the closer hydrogen atom swings away. As discussed in Ref.~\cite{edza+24jcp}, this systematic pivoting correlates with the partial charge distributions, which create pronounced electrostatic gradients across the binding sites. Under the highly symmetric steric confinement of site III, the \ce{NH2}- and \ce{OH}-functionalized variants display more pronounced rotations of 73.92$^\circ$ and 86.29$^\circ$, whereas Zn-MOF-5-\ce{NO2} and Mg-MOF-5-\ce{NO2} prevent any deviation from initial coordination.

When combining metal exchange with ligand functionalization, the large-pore site (I) and the Mg-substituted series systematically follow the established trends of their respective references. However, the Cd-based series deviates sharply. At site II, additional functionalization with \ce{NH2} or \ce{OH} rotates the molecule by more than $80^\circ$, forcing an orthogonal orientation towards the functional group. In the presence of \ce{NO2} termination across all three metal nodes, the rotation at site II stabilizes at $\theta \approx 65^\circ$, aligning the \ce{H2} molecular axis with the electronegative nitro group. Most notably,  in the Cd-based framework, embedding the hydrogen molecule within three \ce{NO2} groups (site III) triggers an exceptionally large rotation of 89.16$^\circ$. Unlike its Zn- and Mg-based counterparts, which lead to zero rotation and push the guest away, the \ce{H2} molecule inside Cd-MOF-5-\ce{NO2} tilts completely, pointing toward the oxygen atoms of the nitro ligands, further supporting the anomalous bond stretching for this system reported in Fig.~\ref{fig:structural_properties}a.


Finally, we evaluate the host–guest interaction distance, quantified as the shortest separation between an H atom of the adsorbate and the nearest framework atom (Fig.~\ref{fig:structural_properties}d). Relative to the initial input distance of 3.00~\AA{}, the resulting relaxed values for the pristine MOF-5 are 2.97~\AA{}, 3.03~\AA{}, and 3.06~\AA{} at position I, II, and III, respectively. These results are in excellent agreement with quantum-chemical calculations on benzoate cluster models, reporting 2.92~\AA{} and 2.99~\AA{} for sites I and II, respectively~\cite{sill+09jacs}. Furthermore, our R2SCAN-D3 values provide a more accurate description of these weak interactions compared to earlier DFT simulations within the local density approximation (LDA), delivering significantly shorter host–guest distances ($\sim$2.6~\AA{})~\cite{cabr+08prb} due to the well-known tendency of uncorrected LDA to overestimate weak interactions.

Isoelectronic metal substitution directly modulates this distance by altering the spatial constraints and the orientation of the guest molecule. Substituting Zn with Mg systematically compresses the host–guest distance within the small-pore environments (sites II and III). As demonstrated in the rotational analysis (Fig.~\ref{fig:structural_properties}c), the \ce{H2} molecular axis tilts toward the carboxylate oxygen atoms in the Mg framework, drawing one of its hydrogen atoms closer to the framework. Conversely, the Cd-substituted variants consistently exhibit larger adsorption distances relative to their Mg counterparts across all three sites. This expansion is driven by the larger steric hindrance of the $\text{Cd}^{2+}$ cation and its distinct rotational profiles, which orient the molecular axis such that the hydrogen atoms are prevented from approaching the cluster.

Ligand functionalization exerts negligible influence at the large-pore site (I) but introduces dramatic, substituent-specific modifications within the small cages. At site II, where a single linker is modified, functionalization universally shortens host–guest separation compared to the unfunctionalized reference. This compression stems from a combination of the localized molecular rotations detailed above and the strong electrostatic gradients generated by the highly negative partial charges of the functional groups~\cite{edza+24jcp}. This trend intensifies significantly under the multi-directional confinement of site III, where the electron-donating \ce{OH} and \ce{NH2} groups drive the closest guest approaches, whereas \ce{OH}-functionalization contracts the interaction distance down to 2.36~\AA{}. In sharp contrast, the bulky and strongly electronegative nitro group induces severe steric and electrostatic repulsion in the Zn- and Mg-based frameworks, pushing away the \ce{H2} molecule (3.23~\AA{}).

When combining metal node exchange with ligand functionalization, the large-pore site (I) and the Mg-substituted series systematically replicate the geometric trends of their respective single-modification baselines. Within the small pore (site II), however, the large coordination sphere of the Cd node introduces a distinct bifurcation when paired with \ce{NH2} and \ce{OH} BDC termination. Because these functional groups force the \ce{H2} molecule into a near-orthogonal orientation relative to the ligand axis, the host-guest distance is highly sensitive to the composition of the substituent. The Cd-\ce{NH2} coupling attracts the dihydrogen to 2.46~\AA{}, whereas the Cd-\ce{OH} variant repels it to 2.68~\AA{}. 
The most remarkable interplay between metal node and functional group occurs at site III for the Cd-MOF-5-\ce{NO2} system. While the \ce{NO2} group acts as a severe steric barrier in Zn and Mg frameworks, pushing the guest molecule far away ($>3.20$~\AA{}), the highly polarizable Cd node completely flips this interaction behavior, attracting the guest molecule to the shortest equilibrium distance of 2.45~\AA{}. 


\subsection{Adsorption Properties}

The energetic stability of \ce{H2} adsorption on the considered MOF-5 derivatives is quantified by the adsorption energy, defined as:
\begin{equation}\label{eq:ads_energy}
\Delta E_\mathrm{ads} = E_\mathrm{MOF,H_2} - \left(E_\mathrm{MOF} + E_\mathrm{H_2}\right),
\end{equation}
where $E_\mathrm{MOF,H_2}$ is the total energy of the relaxed host-guest system, $E_\mathrm{MOF}$ is the energy of the fully optimized empty framework in the chemically modified configuration, and $E_\mathrm{H_2}$ represents the energy of the gas-phase dihydrogen molecule. This definition allows us to evaluate the adsorption ability of each proposed derivative on equal footing, regardless of their relative stability compared to the pristine MOF-5 parent structure.

\begin{figure}[h!]
\begin{center}
\includegraphics[width=\textwidth]{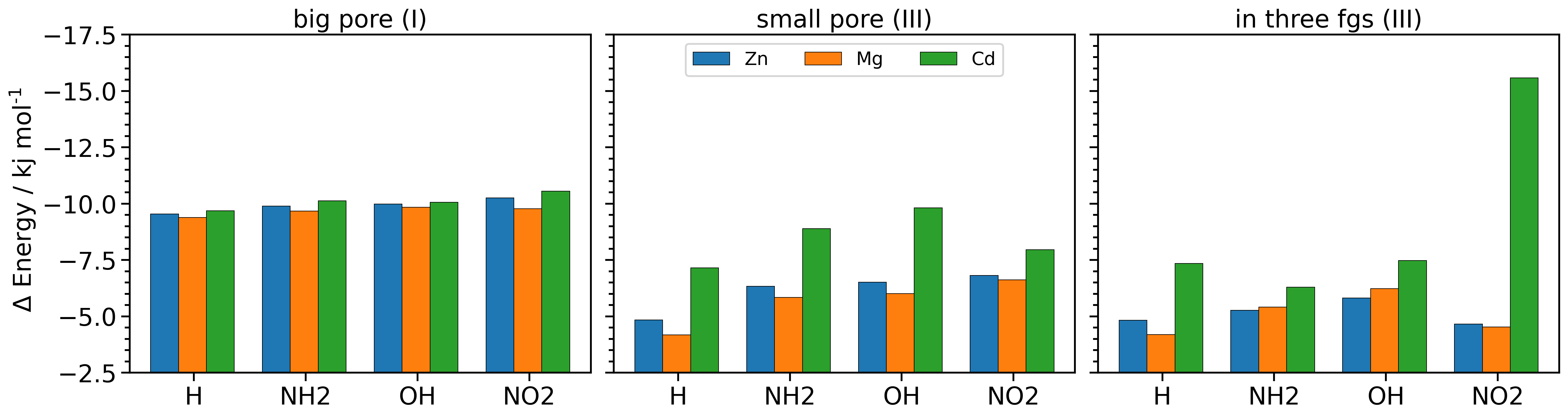}
\caption{\label{fig:adsorption_energies} Calculated adsorption energies ($\Delta E_{\mathrm{ads}}$) of the \ce{H2} molecule within the M-MOF-5-X derivatives with M=\{Zn, Mg, Cd\} and X=\{H, \ce{NH2}, OH, \ce{NO2}\} across three adsorption sites: (I) large pore (II) small pore and (III) small-pore site enclosed by three functional groups.
}
\end{center}
\end{figure}

As shown in Fig.~\ref{fig:adsorption_energies} (further details in Table~S5), all metal-node substitutions and ligand functionalizations lead to negative adsorption energies across all three investigated sites, confirming that every configuration represents a stable physisorption state. For pristine Zn-MOF-5, adsorption energies range from -9.54~\kjmol~(-98.86~meV) for the hydrogen molecule positioned in the large pore (I), to -4.85~\kjmol~(-50.23~meV) and -4.82~\kjmol~(-50.04~meV) in the small pores II and III, respectively. These results follow the same trends of earlier semi-local DFT calculations~\cite{muel+05jpc} (9.5~meV and 21.7~meV obtained for adsorption on sites I and II, respectively) and quantum-chemical estimates on benzoate models~\cite{sill+09jacs} (-7.6~\kjmol~and -4.4~\kjmol for site I and II, respectively). The quantitative discrepancies with our results stem from the known inability of uncorrected, semi-local DFT functionals to account for dispersive interactions, which play a crucial role in describing the absorption process. Our R2SCAN-D3 approach effectively resolves this limitation. Furthermore, local molecular orientation significantly influences the binding energy; enforcing an unrotated configuration at the large-pore site (I) yields a less exothermic adsorption energy of -7.51~kJ~mol$^{-1}$ (-77.85~meV), confirming that the perpendicular reorientation discussed in Fig.~\ref{fig:structural_properties} represents the actual global energy minimum.

Isoelectronic substitution of the metal node yields negligible variations at the large-pore site (I), which consistently maintains the highest absolute adsorption energy across the series. However, metal exchange triggers substantial differences within the confined environments of sites II and III. Substituting Zn with Mg systematically reduces the interaction energy, while Cd enhances adsorption. This behavior matches the relative electronegativities of the metal cations~\cite{cord+08dt}. As established in the structural analysis (Fig.~\ref{fig:structural_properties}d), the host-guest distance is significantly compressed in the Cd-based MOF-5 variants, bringing the adsorbate into closer proximity with the heavy metal-oxo cluster and enhancing the host-guest coupling. Across all metal nodes, the rotated molecular configurations consistently represent the lowest-energy minima.

Linker functionalization introduces significant changes at sites II and III, while leading only to small adjustments at the large-pore site (I). At the small-pore site (II), the absolute adsorption energies increase to -6.34~\kjmol~(-65.68~meV), -6.52~\kjmol~(-67.59~meV), and -6.81~\kjmol~(-70.61~meV)~for \ce{NH2}, OH, and \ce{NO2} BDC termiations, respectively, corresponding to a net stabilization of up to 1.97~\kjmol~(20.38~meV) relative to the pristine baseline. This enhancement is driven by the localized negative partial charges on the functional groups~\cite{edza+24jcp}, which amplify electrostatic interactions with the guest molecule. This effect is reinforced by the localized and closer distance of \ce{H2} toward the substituent. Interestingly, adsorption at site III further enhances the binding energy compared to pristine MOF-5, yielding values of -5.27~\kjmol~(-54.67~meV), -5.82~\kjmol~(-60.27~meV), and -4.67~\kjmol~(-48.35~meV)~for \ce{NH2}-, OH-, and \ce{NO2}-functionalization, respectively. These results concurrently suggest that excessive steric crowding begins to penalize the binding ability. The nitro group attached to Zn-MOF-5 at site III represents the exception to this trend. Due to the large hindrance of the closely located \ce{NO2} groups and their mutual repulsion, the adsorption energy for \ce{H2} is even lower than the pristine benchmark (-4.67~kJ~mol$^{-1}$, corresponding to -48.35~meV).

Ultimately, the combination of metal-node substitution and ligand functionalisation is the key mechanism for optimizing host-guest interactions. While the large-pore site (I) is not influenced by BDC modifications, the small-pore environments reveal clear cooperative effects. Cd-substitution outperforms all other variants when paired with any functional group. While bulky functional groups act to narrow the pore volume near the oxo-cluster, the large atomic radius of the Cd node counteracts this constraint by physically opening the local geometry, optimizing the spatial accommodation of the guest near the cluster face. Conversely, the Mg series exhibits the weakest binding affinities at site II, though it recovers slightly at site III. The most outstanding performance is achieved in the Cd-MOF-5-\ce{NO2} framework at site III, where our calculations predict an exceptionally high adsorption energy of -15.58~\kjmol~(-161.53~meV). This value is more than three times larger than the value obtained for pristine MOF-5, completely reversing the steric penalties obtained when functionalizing the Zn- and Mg-based frameworks with three nitro groups. 

To put this performance into practical perspective,  thermodynamic models for ambient-temperature hydrogen storage target an optimal adsorption enthalpy window of 15--25~\kjmol~to achieve high deliverable capacities under practical pressure-swing conditions~\cite{sutt+24cej}. While pristine MOF-5 falls far short of this target, scoring -4.8~\kjmol{} at site III, this $>3$-fold energetic enhancement positions Cd-MOF-5-\ce{NO2} squarely within the lower threshold required for near-ambient operation.
This extraordinary energetic stabilization validates the structural anomalies discussed in Fig.~\ref{fig:structural_properties} and confirmed by charge-density difference analysis (Fig.~S10). Importantly, while this interaction is substantially enhanced compared to pristine MOF-5, it remains well within the window of reversible physisorption~\cite{ma+24ijhe,suh+12cr}. Operating via non-covalent polarization, this binding regime allows hydrogen to be repeatedly adsorbed and desorbed without structural fatigue or chemical degradation of the host framework~\cite{suh+12cr, meek+24natm}. It is worth noting that the physisorption regime disclosed here prevents the evaluation of recovery times based on models developed on planar surfaces hosting localized chemical desorption~\cite{zhan-chen25drm}.


\begin{figure}[htbp]
\begin{center}
\includegraphics[width=\textwidth]{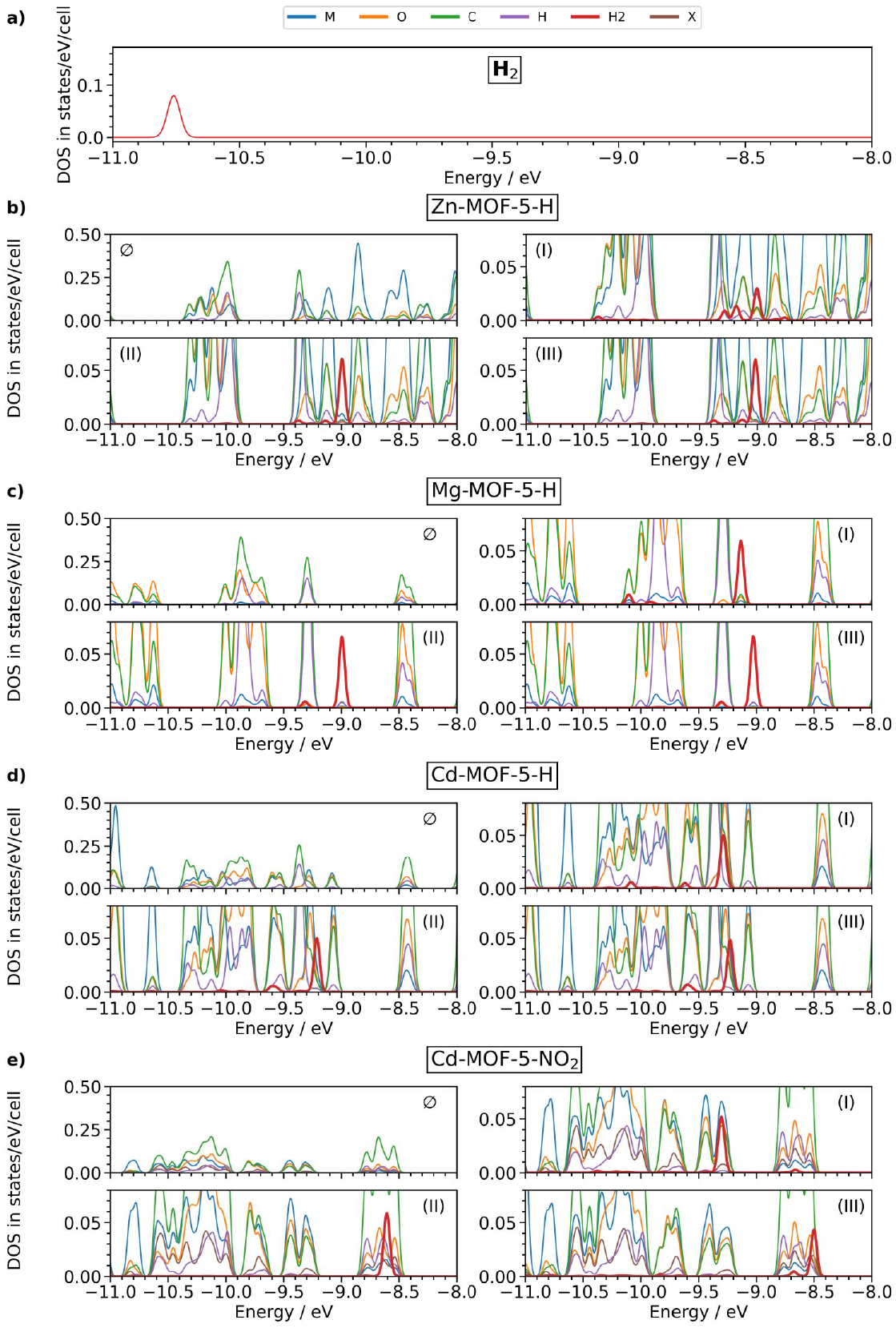}
\caption{\label{fig:pdos} Projected densities of states (pDOS) for a) an isolated \ce{H2} molecule, b) pristine Zn-MOF-5, c) Mg-MOF-5-H, d) Cd-MOF-5-H, and e) Cd-MOF-5-\ce{NO2}, comparing the results obtained for the empty framework ($\emptyset$) against \ce{H2} adsorption in the large pore (I), the small pore (II), and the small pore enclosed by three functional groups (III). The contribution of the \ce{H2} molecule is highlighted by the thick red curve in each plot. All energies are offset to the vacuum level set to zero.}
\end{center}
\end{figure}

To gain insight into the quantum-mechanical mechanisms ruling host-guest interactions in the modified MOF-5, we examine the pDOS of selected systems, including Mg- and Cd-substituted MOF-5 and the most stable Cd-MOF-5-\ce{NO2} derivative. The pDOS for the remaining structures in the dataset are reported in Figures~S2--S9. As a reference, we report in Fig.~\ref{fig:pdos}a the DFT spectrum of the isolated, gas-phase \ce{H2} molecule, characterized by a distinct peak at -10.75~eV, corresponding to the $\sigma$-state. Although this energy value underestimates the experimental reference of -15.4~eV~\cite{blea32pr,herz-jung72jms} for the ionization potential, it aligns well with earlier DFT predictions~\cite{grit+16jcp}, which notoriously underestimate absolute electronic energies due to self-interaction errors. 

Embedding the hydrogen molecule within pristine MOF-5 (Zn-MOF-5-H) results in an energy increase of the $\sigma$-band of \ce{H2} by about 1.75~eV, with the peak appearing around -9~eV in all adsorption configurations (Fig.~\ref{fig:pdos}b). While MOF-5 alone does not exhibit any electronic states at that energy, the introduction of \ce{H2} triggers site-specific hybridization pathways. Specifically, the placement of the molecule in the large pore (I) leads to interactions with Zn $d$-orbitals as well as carbon and oxygen $p$-states. This orbital mixing induces noticeable crystal field splitting in the metal $d$-manifold, consequently strengthening adsorption (compare Fig.~\ref{fig:adsorption_energies}). Conversely, the pDOS profiles of the small-pore sites (II and III) mirror each other closely, reflecting their nearly identical adsorption energies and confirming a weaker interaction compared to site I. In both configurations, the $\sigma$-state of \ce{H2} hybridizes predominantly with the metal node, inducing only minor modifications to the spectrum of the Zn $d$-orbitals.

Similarly, in Mg-MOF-5-H, the $\sigma$-band of the adsorbate shifts toward -9~eV. However, the large electronic gap between -9.2 and -8.5~eV in the pDOS of the empty cage substantially inhibits hybridization, except for residual couplings with the hydrogen atoms in the BDC linker (Fig.~\ref{fig:pdos}c). Minor electronic interactions with the framework occur in the large-pore site (I) at deeper valence energies (-10.1~eV), involving localized carbon states. In contrast, the small-pore configurations (II and III) exhibit very similar pDOS profiles. Here, weak hybridization occurs at -9.6~eV, where mixed Mg, C, and O bands reside.

Cd-MOF-5-H, which scores the highest absolute adsorption energy among the H-passivated series, displays a distinct electronic response (Fig.~\ref{fig:pdos}d). Here, the $\sigma$-state of \ce{H2} undergoes relative stabilization, shifting to lower energies (-9.25~eV) compared to the Zn- and Mg-based analogs. At the large-pore site (I), the overlap with the Cd $4d$ bands and the $2p$ shells of the surrounding oxygen and carbon atoms is maximized compared to the more confined small-pore environments (sites II and III), explaining why the Cd framework maintains superior binding properties relative to Zn and Mg substitutions across all configurations. This extended interaction is macroscopically evidenced by widespread split satellite peaks in the deeper valence bands, indicating generalized orbital re-hybridization.

The most interesting electronic behavior is displayed by Cd-MOF-5-\ce{NO2}, where the electronic states of the guest molecule interact most strongly with those of the hosting environment, exhibiting significant differences depending on the adsorption site (Fig.~\ref{fig:pdos}e). Inside the large pore (I), the $\sigma$-band appears at -9.3~eV and overlaps almost completely with the Cd $4d$-states. In the small-pore sites, the occupied \ce{H2} state appears at higher energies, namely at -8.6~eV in site II and at -8.5~eV in site III. Here, hybridization with C states, as well as with residual Cd, O, and H orbitals, is most pronounced, suggesting robust electronic interaction between the molecule and the framework. This finding is consistent with the significantly higher absolute adsorption energies obtained not only relative to the H-passivated counterpart but also to all the other variants in the series (Fig.~\ref{fig:adsorption_energies}).

\section{Summary and Conclusions}
In summary, we presented a comprehensive first-principles investigation of the hydrogen adsorption properties of MOF-5 and its derivatives obtained through systematic metal-node substitution and linker functionalisation. Adopting a state-of-the-art DFT approach, embedded in an efficient automated workflow implemented in the \texttt{Python} library \texttt{aim$^2$dat}~\cite{sass+26es}, we systematically evaluated a configurational space of 36 distinct host-guest configurations. Our findings reveal the fundamental, intrinsic ground-state electronic mechanisms, localized charge polarization, and intrinsic binding strengths. Specifically, they indicate that, while the large-pore site is substantially insensitive to structural and electronic variations of the linkers, the sterically confined small-pore environments can be fine-tuned by both metal substitution and chemical modifications of the ligands. Cationic exchange with the larger, highly polarizable Cd$^{2+}$ node consistently outperforms Zn and Mg. Linker functionalization introduces pronounced localized electrostatic gradients, which are further amplified by the introduction of \ce{NO2} groups attached to the BDC molecules. The resulting Cd-MOF-5-\ce{NO2} system scores an exceptional adsorption energy of -15.58~kJ~mol$^{-1}$ (-161.53~meV) in the small pore, which is more than three times larger than the value achieved at the same site by the pristine reference. This massive energetic gain is mirrored by a drastic reduction of the host-guest distance down to 2.45~\AA{}, a complete perpendicular reorientation of the \ce{H2} molecular axis, and an intramolecular H–H bond stretching of $\sim$0.01~\AA{}. This behavior is driven by a maximized electronic overlap of the $\sigma$-orbital with the Cd $4d$ bands and the $2p$ valence states of the \ce{NO2} oxygen atoms. 

These findings establish Cd-MOF-5-\ce{NO2} as the best conceptual candidate to enhance the \ce{H2} adsorption ability of MOF-5 architectures. While the toxicity of cadmium and \ce{NO2} hinders the synthesis of this specific compound for safety reasons, this computational study establishes a feasible route to include MOF-5 among effective framework materials for hydrogen storage. Following the strategy successfully established in previous work~\cite{sass+24ic}, future research can leverage this rationale to identify equally efficient, non-toxic surrogates. 
Looking forward, upcoming studies based on Monte Carlo simulations are expected to shed light on the thermodynamic and kinetic effects that are crucial for a quantitative assessment of sorption cycling under operational conditions.



\funding{
This work was financed by the German Federal Ministry of Education and Research (Professorinnenprogramm III) and by the State of Lower Saxony (Professorinnen f\"ur Niedersachsen and ``Grundlagen f\"ur Photovoltaik-Technologies der Zukunft'' -- FuturePV). J.E. appreciates financial support from the Nagelschneider Stiftung. H.-D. S. acknowledges funding from the German Research Foundation (Deutsche Forschungsgemeinschaft, DFG) via a Walter-Benjamin Fellowship (project No. 561059489). Computational resources were provided by the North-German Supercomputing Alliance (NHR), project nic00084.
}

\roles{\textbf{J.E.} Conceptualization, Methodology, Investigation, Writing – original draft; \textbf{H.-D.S.} Software, Methodology, Writing – original draft; \textbf{C.C.} Conceptualization, Funding acquisition, Resources, Supervision, Writing – review \& editing}

\data{
The data produced in this study are openly available in Zenodo at DOI: \url{https://10.5281/zenodo.21937751}.
}

\suppdata{
Additional information regarding the computational workflow, as well as structural, adsorption, and electronic properties of the considered frameworks is provided in the Supplementary Information.
}
\bibliographystyle{iopart-num}
\providecommand{\newblock}{}

\end{document}